\documentclass{llncs}

\usepackage{lineno,hyperref}
\usepackage{xcolor}
\usepackage{multirow}
\usepackage{booktabs}
\usepackage{rotating}
\usepackage{todonotes}
\usepackage{amsmath,amssymb}

\begin{document}

\begin{frontmatter}

% Methodological issues resulting in near-perfect results on imbalanced datasets: a review of literature on the TPEHGDB dataset the imbalanced TPEHGDB dataset: a review of recent literature

\title{Overly Optimistic Prediction Results on Imbalanced Data: a Case Study of Flaws and Benefits when Applying Over-sampling}

% Achieving Overly Optimistic Results in Predictions Concerning Minority Groups: a Methodological Oversampling Flaw

\author{Gilles Vandewiele\inst{1} \and
	Isabelle Dehaene\inst{2} \and
	Gy\"{o}rgy Kov\'{a}cs\inst{4} \and
	Lucas Sterckx\inst{1} \and
	Olivier Janssens\inst{1} \and
	Femke Ongenae\inst{1} \and
	Femke De Backere\inst{1} \and
	Filip De Turck\inst{1} \and
	Kristien Roelens\inst{2} \and
	Johan Decruyenaere\inst{3} \and
	Sofie Van Hoecke\inst{1} \and
	Thomas Demeester\inst{1}}

\institute{IDLab, Ghent University -- imec \\ Technologiepark-Zwijnaarde 126, Ghent, Belgium \and
	Department of Gynaecology and Obstetrics, Ghent University Hospital \\ Corneel Heymanslaan 10, Ghent, Belgium \and
	Department of Intensive Care Medicine, Ghent University Hospital \\ Corneel Heymanslaan 10, Ghent, Belgium \\
	\email{\{firstname\}.\{lastname\}@ugent.be} \and
	Analytical Minds Ltd. \\ Arpad street 5, Beregsurany, Hungary\\
	\email{gyuriofkovacs@gmail.com}}

\maketitle

\begin{abstract}
%169 words
Information extracted from electrohysterography recordings could potentially prove to be an interesting additional source of information to estimate the risk on preterm birth. Recently, a large number of studies have reported near-perfect results to distinguish between recordings of patients that will deliver term or preterm using a public resource, called the Term/Preterm Electrohysterogram database. However, we argue that these results are overly optimistic due to a methodological flaw being made. In this work, we focus on one specific type of methodological flaw: applying over-sampling before partitioning the data into mutually exclusive training and testing sets. We show how this causes the results to be biased using two artificial datasets and reproduce results of studies in which this flaw was identified. Moreover, we evaluate the actual impact of over-sampling on predictive performance, when applied prior to data partitioning, using the same methodologies of related studies, to provide a realistic view of these methodologies' generalization capabilities. We make our research reproducible by providing all the code under an open license.
\end{abstract}

\keywords{
preterm birth risk estimation \and over-sampling \and electrohysterography
}

\end{frontmatter}

\section{Introduction}
Giving birth before 37 weeks of pregnancy, which is referred to as preterm birth, has a significant negative impact on the expected outcome of the neonate. According to the World Health Organization (WHO), preterm birth is one of the leading causes of death among young children, and its prevalence ranges from 5\% to 18\% globally~\cite{liu2016global}. As preterm labor is currently not yet fully understood, a gynecologist experiences difficulties to assess whether a patient presenting with symptoms suggestive for preterm labor, will actually deliver preterm or not. In order to support these experts in their assessment, several studies have already investigated the added value of a predictive model~\cite{meertens2018prediction,watson2017quipp,de2017timing,garcia2017can,vandewiele2019time}. These models are based on a large number of variables extracted from clinical sources, such as the electronic health record, including the gestational age, results of a biomarker, cervical length, clinical history. \\

One interesting alternative that could be used as an additional source of valuable information for preterm birth risk prediction is electrohysterography (EHG). EHG is a technique that measures the electronic potentials of the uterine muscle by attaching patches around the abdomen of the woman. The technique can be seen as a new and better alternative to measuring uterine activity. As opposed to the intrauterine pressure catheter (IUPC), it is non-invasive and has no infection, perforation or hemorrhage risk~\cite{euliano2013monitoring}. Moreover, EHG possesses a higher signal-to-noise ratio than tocography, especially in certain groups, such as obese women~\cite{euliano2007monitoring,davies2010obesity}. \\

This paper analyzes a large set of studies providing results using a public resource, further detailed in Section~\ref{sec:relwork}, called the `Term/Preterm ElectroHysteroGram DataBase' (TPEHGDB)~ \cite{fele2008comparison}. While the problem of predicting preterm birth is far from solved in reality, many of these studies report near-perfect prediction results. After carefully considering the presented machine learning methodologies, we argue that specific subtle, yet critical, choices in the data pre-processing setup may lead to information leakage from the held-out evaluation set into the training set. In particular, we often observed the incorrect application of over-sampling techniques to circumvent data imbalance issues~\cite{he2009imbalanced}, which often arise in medical applications as the number of healthy (negative) cases often greatly exceeds the number of cases with a certain disorder or disease (positive). As a consequence, the evaluation results no longer represent an evaluation on actually unseen data. Other studies have already discussed, on a conceptual level, the danger of applying over-sampling before data partitioning~\cite{santos2018cross,lusa2015joint}. In those studies, these dangers have been demonstrated on several benchmark datasets from the UCI Repository~\cite{asuncion2007uci}. In this work, we focus on the studies using that same dataset, which we suspect of incorrect data handling. In contrast to prior studies, we reproduce each of the suspected studies to provide empirical evidence that over-sampling was applied before data partitioning \\

As one of the reviewers of this work noted, the methodological errors pointed out in this work are quite obvious. However, the main value in analyzing these errors in the underlying work, comes from the fact that so many studies were eventually published in high-profile journals with an obvious flaw not detected despite rigorous reviewing. We aim to raise awareness, but we do not intend to point the finger at any individual study or methodology. However, in order to provide clear insights into the potential pitfalls in data pre-processing and model evaluation, we try to divide the referred works according to different potential issues, to the best of our ability (see Section~\ref{sec:relwork}). After a short discussion on the use of  over-sampling techniques (Section~\ref{sec:oversampling}), the remainder of the paper (Section~\ref{sec:results}) is devoted to reporting our own results on the TPEHGDB dataset for the various models, engineered features, and over-sampling techniques available in literature. Moreover, we investigate the actual impact of applying over-sampling, when performed correctly, and the added value of tuning the over-sampling algorithm and its corresponding hyper-parameters.  \\

The contributions of this paper are summarized as follows:
\begin{itemize}
	\item We provide an exhaustive overview of studies using the TPEHGDB dataset of EHG recordings which tackle the challenge of preterm birth risk estimation. We elaborate on three types of potential issues in the evaluation methodology of studies reporting near-perfect prediction results. We then focus on one specific issue, i.e. applying over-sampling before partitioning the data into a mutually exclusive training and testing set.
	\item We explore the individual quality of each of the features discussed in these studies by measuring their capability to distinguish between term and preterm recordings.
	\item We experimentally point out the consequences of incorrectly applying over-sampling (i.e., before data partitioning).
	\item We discuss the actual impact of existing over-sampling techniques to counteract data imbalance, when applied prior to data partitioning, using the same methodologies as related studies in order to provide a realistic view of the performance of their proposed methodologies. Moreover, we study whether tuning the over-sampling algorithm and its corresponding hyper-parameters can result in an increase in the predictive performance.
\end{itemize}

This work extends Vandewiele et al.~\cite{vandewiele2019critical}, where we first pointed out the potential issues of the studies reporting near-perfect results on the TPEHGDB dataset. In our previous work, we discussed these potential issues to then reproduce only one of the studies that we suspected of applying over-sampling before data partitioning. The reason for reproducing that specific study was that it only used features that were already provided by the owners of the TPEHGDB dataset. In contrast, we reproduce eleven suspected studies in this study to then compare the predictive performance when sampling is applied before partitioning to the performance when sampling is applied after partitioning. As such, we implemented over fifty different features that were discussed in those works. In addition to reproducing these studies, we also study the capability of each of those features individually to discriminate between term and preterm records and assess the impact of algorithm selection and hyper-parameter tuning of over-sampling techniques. We provide the source code for the feature extraction and experiments to serve as a basis for future research endeavours\footnote{\url{https://github.com/GillesVandewiele/EHG-oversampling/}}.

\section{Studies estimating preterm birth risk using the TPEHGDB}\label{sec:relwork}

\begin{figure}[t!]
	\centering
	\includegraphics[scale=0.6]{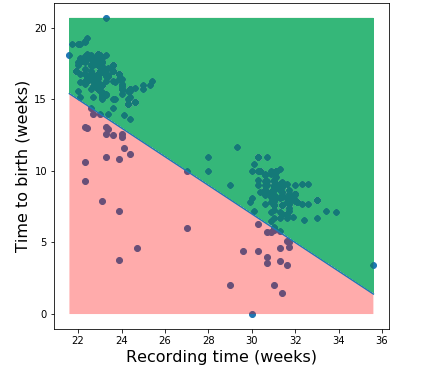}
	\caption{The number of weeks till birth expressed as a function of the gestational age in weeks at the time of recording. All data points within the red area correspond to preterm deliveries, while the ones within the green area correspond to term deliveries.}
	\label{fig:eda1}
\end{figure}

In 2008, a public dataset called TPEHGDB (Term/Preterm ElectroHysteroGram DataBase), containing 300 records that correspond to 300 different patients and pregnancies, was released on PhysioNet~\cite{fele2008comparison,PhysioNet}. Each record consists of three raw bipolar signals that express the difference in electric potentials, measured by four electrodes placed on the abdomen. In addition, each record is accompanied by clinical variables, such as the gestational age at recording time, the age and weight of the mother, and whether or not an abortion had occurred in the patient's medical history. The recordings can be categorized as being captured at an earlier stage in pregnancy (gestational age of $23.11\pm0.77$ weeks) or at a later stage  ($31.09\pm1.05$ weeks). Recordings were captured at a frequency of 20~Hz for about 30 minutes. In Figure~\ref{fig:eda1} the number of weeks till birth is shown as a function of the gestational age at the time of recording and displayed according to term or preterm delivery. A clear imbalance is present in the dataset with significantly more term deliveries (262 in total, green area) than preterm ones (38, red area). \\

We now review machine learning studies using the TPEHGDB dataset. Screening for the term `machine learning', we selected a subset of 82 studies from the 160 citations, present in Google Scholar on January 2020, to the original paper. These studies were then manually double-checked on whether they were a machine learning study and presented clear prediction results on the TPEHGDB dataset. In total 24 studies were identified using machine learning on the TPEGHDB. While many of these studies are reporting near-perfect results, these should be interpreted very cautiously as a flaw may be present in their methodology, leading to biased and overly optimistic reported metrics. \\

We can categorize these flaws into three different categories. First, studies often apply cross-validation on a subset of data subsampled from the original dataset. Performing this kind of preprocessing, in a machine learning context, without any kind of argumentation, raises doubts as it drastically increases the variance of the obtained results and avoids the problem of imbalanced data, which does not reflect reality in terms of potential applications~\cite{baghamoradi2011evaluation,naeem2014new,naeem2013kl,sim2014evaluation,far2015prediction,beiranvand2017investigating,sadi2017relevant,subramaniam2018classification}. Second, there are a few studies that extract segments from the EHG signals and use these for classification. When doing this, it is very important to ensure that segments extracted from the same original signal are not divided into both training and testing set~\cite{despotovic2018machine,shahrdad2018detection}. Finally, there are many studies applying over-sampling before partitioning the data into two mutually exclusive sets in order to make the distribution of classes more uniform~\cite{fergus2013prediction,ren2015improved,hussain2015dynamic,idowu2015artificial,ahmed2016multivariate,fergus2016advanced,acharya2017automated,jager2018characterization,hoseinzadeh2018use,khan2019characterization,peng2019evaluation}. In this work, we focus on reproducing the results from the final category of flaws. \\

At the time of writing (January 2020), out of the 160 citations to the original paper which introduced the TPEHGDB dataset, we have found three machine learning studies that were accessible, tackled preterm birth risk estimation and, to the best of our knowledge, had a sound evaluation methodology~\cite{ryu2015time,janjarasjitt2017examination,sadi2015contraction}. In the study of Sadi-Ahmed et al.~\cite{sadi2015contraction}, all records taken before 26 weeks of gestation were filtered from the dataset, resulting in a dataset of 138 recordings taken after the 26th week of gestation. All of these signals were processed in order to detect contractions through Auto-Regressive Moving Averages (ARMA). From the detected contractions, features were extracted such as the total number of contractions, average duration and average time between contractions. An accuracy score of 0.89 to distinguish between term and preterm pregnancies was achieved within this study. This makes it hard to assess the clinical use of such a model, since an accuracy score of 0.86 can be achieved by always predicting term birth on this filtered dataset. Janjarasjitt et al.~\cite{janjarasjitt2017examination} proposed a feature type based on a wavelet decomposition of the signals. This feature was evaluated by tuning a threshold on a single feature in a leave-one-out cross-validation scheme. A sensitivity and specificity of 0.6842 and 0.7133 were achieved. While these scores are promising, they might be rather optimistic due to the fact that the evaluation happened in a leave-one-out scheme. As such, the performance of the sample entropy feature, provided along with the original data, closely matches, and sometimes even outperforms, that of the proposed feature. Nevertheless, the wavelet-based feature may be an interesting and complementary addition to the feature set. In the work of Ryu et al.~\cite{ryu2015time} a similar study was performed in which they proposed a feature based on Multivariate Empirical Mode Decomposition (MEMD). They evaluated the added value of their feature, by subsampling a balanced dataset of 38 term and 38 preterm records, 100 times, from the original dataset. They found that the AUC improved from 0.5698 to 0.6049 by adding their feature to the dataset. While this subsampling strategy again avoids the problem of imbalanced data, which is reflected in the original dataset, it does show an improvement in AUC and thus indicates that adding the MEMD-based feature to the dataset could be beneficial for the predictive performance. Moreover, due to the many repetitions of the experiment, the sample mean better reflects the real mean. 

\section{Imbalanced learning and the impact of over-sampling prior to data partitioning}\label{sec:oversampling}

As illustrated in Figure \ref{fig:eda1}, the TPEHGDB dataset is highly imbalanced. An imbalance in the number of samples can make the majority class overly represented in the loss function, leading to a poor generalization of the minority class. In order to tackle this, many authors apply \emph{imbalanced learning}~\cite{he2009imbalanced} techniques to improve the performance of TPEHGDB classification. One of the most popular imbalanced learning approach is \emph{over-sampling} which generates artificial training samples for the minority class by making relatively mild assumptions on the local distribution of the data. The popularity of over-sampling can be illustrated by the numerous variants and successful applications summarized in a recent review \cite{chawla2018smote}, and can be attributed to the fact that it is agnostic to the model being used since it is a preprocessing technique. Although over-sampling techniques are popular and easy to use, there are many pitfalls to avoid when they are applied. Special care must be taken when over-sampling is used in cross-validation to avoid \emph{information leakage}. As the sample generation is based on the entire dataset, the generated artificial samples contain information about many (in some cases all) original samples. Thus, carrying out over-sampling \emph{before} splitting into training and testing sets might leak information from the original testing samples to the artificially generated training samples, leading to overly optimistic validation scores. It is therefore of key importance to carry out the over-sampling \emph{after} selecting a training and testing set. \\

Further, we emphasize that the correct number of artificial instances to be added depends on the distribution of the data and the subsequent data processing pipeline and should thus be tuned. For example, the performance of local classifiers (like k-Nearest Neighbors) clearly depends on the density of samples near the decision boundary, if the density of positive and negative samples differs, the classifier will be biased towards the class with more samples in a unit volume~\cite{he2009imbalanced}. However, the local density of positive and negative samples near the decision boundary is independent from the total number of samples. Even if a dataset is severely imbalanced, kNN can operate ideally if the density of points near the decision boundary is approximately equal. Similarly, balancing a dataset without checking the local densities near the decision boundary can lead to highly imbalanced local densities near the decision boundary, thus, deteriorating the performance of the classifier.\\ 

\begin{figure}
	\centering
	\includegraphics[width=\linewidth]{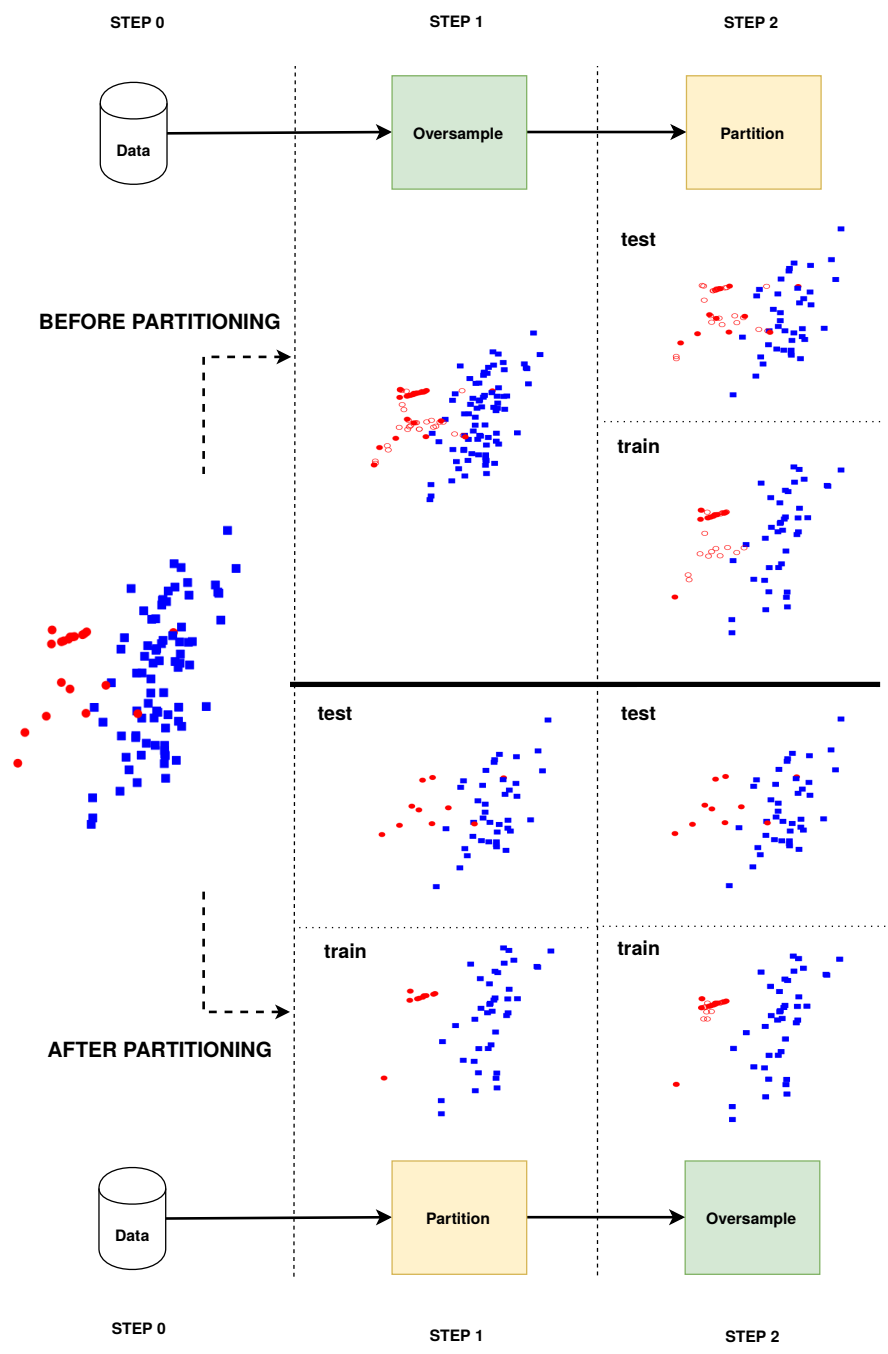}
	\caption{Comparing the impact of applying over-sampling prior to data partitioning to applying over-sampling after data partitioning on an artificial two-dimensional classification problem.}
	\label{fig:over-sampling}
\end{figure}

We highlight the impact of applying over-sampling prior to the data partitioning on an artificially generated dataset. We generated a binary classification problem with 100 samples. Twenty samples were marked positive (red circles), and the others negative (blue squares). The generated dataset is depicted on the left of Figure~\ref{fig:over-sampling} (step 0). We now compare the effect of over-sampling data prior to partitioning with the effect of over-sampling after partitioning. In the former approach, we over-sample the data prior to partitioning, which introduces data leakage by generating training samples that are highly correlated with original data points that will end up in the testing set (step 1). Moreover, some of the generated artificial samples will be distributed to the testing set as well (step 2). This causes highly optimistic results that merely reflect the model's capability to memorize samples seen during training, rather than its predictive performance if it were applied in a real-world setting on unseen data. In the latter, we first partition our data into two mutually exclusive sets (step 1). Then, we create artificial samples (red, unfilled circles) that are highly correlated to the training samples of the minority class (step 2) in order to have a similar number of samples for both classes in our training set. \\

\section{Results}\label{sec:results}

In this section, we re-evaluate the related studies mentioned above by assessing the predictive power of the features proposed in those works, and by reproducing the results of their methodology. Afterwards, we study the true impact of over-sampling on the models' predictive performance.

\subsection{EHG Signal Preprocessing}

We used the EHG signals provided by the original authors~\cite{fele2008comparison,PhysioNet}, that were filtered using a Butterworth digital filter with cutoff frequencies 0.08 to 4.0 Hz. The first and last 150 seconds (corresponding to 3,000 measurements) of the recording, which could contain noise caused by the installation or removal of the recording setup, were removed. Two signals were discarded due to the fact that the recording length was shorter than 30 minutes. In total, this results in a dataset of 298 recordings each containing 3 signals of 30,000 measurements.

\subsection{Predictive power of features}

For reproducing the results from the aforementioned studies, we re-implemented all necessary features, provided they were described in sufficient detail to allow for reconstruction. Missing details were filled in as optimistically as possible. If hyper-parameters were missing, we picked those that resulted in an optimal predictive performance, according to a tuning phase. For the study of Ahmed et al.~\cite{ahmed2016multivariate}, the sample entropy was used instead of the discussed fuzzy entropy technique, as the latter was not described clearly. It should further be noted that although it is possible that there are minimal differences between the original studies and our reproduction, it is very unlikely that these differences resulted in actual near-perfect predictive performances when over-sampling would have been applied correctly. \\

In order to evaluate the capability to distinguish term from preterm signals of each feature individually, we measure the AUC by applying 10,000 bootstrapping iterations and report mean and corresponding standard deviation. This is done by varying a threshold over the full range of a feature for each of the resamplings of the entire dataset. We report this for each of the three channels, each corresponding to a bipolar signal, separately using (i)~all samples, (ii)~the samples with early recording times ($\leq$ 26 weeks) and (iii)~the samples with late recording times ($>$ 26 weeks). As there are a few thousand features per signal, we report only the 10 top performing features. As informative features can result in either high or low AUC scores (i.e. when the feature values of the positive samples are lower than the negative samples), we define top performing as the maximal difference between the measured AUC, using all samples, and $0.5$, which resembles random guessing. The AUC scores, their standard deviations and the corresponding channel of these top features can be found in Table~\ref{tab:auc_scores}. For a more detailed explanation on the reported features, and how to extract them, we refer readers to the corresponding related work. A large number of features are extracted from a spectral representation, obtained by applying Empirical Mode Decomposition (EMD) and/or Wavelet Packet Decomposition (WPD). In those cases, we mention how many times the EMD has been applied and mention the level (number of A) and the type of the final coefficient (A or D) of the WPD. Note that this uni-variate analysis only measures the discriminative capability of individual features, and not that of combinations of features. Yet, for brevity, we have chosen to limit ourselves to this first-order analysis. \\ 

From these results, we see that performing EMD and/or WPD often results in more informative features, as many of the top features come from these categories. Moreover, we see that the EHG signal from channel 3 is the most informative one. Nevertheless, the informativeness of the individual features is rather limited, with a maximal AUC ranging roughly between $0.60$ and $0.65$, with high standard deviations, obtained by extracting the Higuchi Fractal Dimension after performing WPD on the output of the EMD.

\begin{table}[]
	\centering
	\scriptsize
	\begin{tabular}{|c|c|c|c||cccc|}
		\toprule Feature & \textsc{emd} & \textsc{wpd} & From & Ch. & All & Early & Late \\ \midrule
		Sample Entropy & 2 & \textsc{aaa} & \cite{acharya2017automated} & 3 & $69.0 \pm 4.2$ & $72.6 \pm 6.8$ & $63.35 \pm 5.6$ \\
		Standard Deviation & 2 & \textsc{aaaa} & \cite{acharya2017automated} & 3 & $36.2 \pm 4.4$ & $36.7 \pm 6.3$ & $36.4 \pm 6.2$ \\
		Teager-Kaiser Energy & 2 & \textsc{aaaa} & \cite{acharya2017automated} & 3 & $35.5 \pm 4.4$ & $34.8 \pm 6.3$ & $36.0 \pm 6.3$ \\
		Interquartile Range & 9 & \textsc{aaaad} & \cite{acharya2017automated} & 1 & $34.4 \pm 4.9$ & $33.0 \pm 6.3$ & $36.0 \pm 8.0$ \\
		Higuchi Fractal Dim. & 3 & \textsc{ad} & \cite{acharya2017automated} & 3 & $67.6 \pm 4.7$ & $62.8 \pm 7.1$ & $73.0 \pm 6.2$ \\
		Sampen (m=4, s=5) & $\times$ & $\times$ & \cite{ahmed2016multivariate} & 3 & $34.0 \pm 4.6$ & $25.2 \pm 5.9$ & $43.4 \pm 7.3$ \\
		1$^{\textrm{st}}$ Yule-Walker coef & 2 & \textsc{a} & \cite{hoseinzadeh2018use} & 3 & $68.4 \pm 4.6$ & $69.5 \pm 7.0$ & $66.2 \pm 6.4$ \\
		Median Frequency & $\times$ & $\times$ & \cite{jager2018characterization} & 3 & $62.7 \pm 4.2$ & $65.5 \pm 5.9$ & $59.2 \pm 6.5$ \\
		Wavelet Log Var Diff & $\times$ & \textsc{aaad} & \cite{janjarasjitt2017evaluation} & 3 & $37.4 \pm 4.2$ & $29.3 \pm 5.6$ & $48.1 \pm 6.6$ \\
		FWL Peak Power & 7 & $\times$ & \cite{sadi2017relevant} & 1 & $34.6 \pm 4.9$ & $36.5 \pm 7.9$ & $32.9 \pm 6.1$ \\ \bottomrule
	\end{tabular}
	\caption{The ability of the reproduced features from channel 1 to distinguish between term and preterm samples, expressed as the average AUC and its corresponding standard deviation, measured using 10,000 bootstrap iterations. When EMD has been applied, we mention the number of iterations. When WPD has been applied, we provide the level (number of A) and the type of the final coefficient (A or D).}
	\label{tab:auc_scores}
\end{table}

\subsection{The impact of over-sampling}

In this section, we demonstrate the impact of applying over-sampling before data partitioning. We first show how near-perfect predictive performance can be achieved on randomly generated data with no predictive signal at all. Afterwards, we reproduce eleven studies, mentioned in Section~\ref{sec:relwork} and measure the predictive performance when over-sampling is applied either before or after data partitioning.

\subsubsection{Artificial dataset experiment}
To exemplify the effect of bias, introduced by over-sampling before partitioning the data, we artificially generate 10,000 five-dimensional points, with each variable sampled from a uniform distribution ($X_i \sim U(0, 1)$ for $i \in \{ 1, \ldots, 5 \}$). We then create an imbalanced binary classification problem by randomly labeling 90\% of the data as negative and the remaining 10\% as positive. As the data is randomly generated, we expect the predictive performance of a classifier to be as good as random guessing. Nevertheless, due to applying the Synthetic Minority over-sampling Technique (SMOTE) before data partitioning, an Area Under the receiver operating characteristic Curve (AUC) of $0.95$ on a `held-out' testing set can be achieved. This in contrast to an AUC of $0.49$ and $0.48$ when applying no SMOTE or SMOTE after data partitioning respectively, which closely resembles random guessing.

\subsubsection{Reproducing studies using TPEHGDB}
In this section, we re-implement the methodologies of studies mentioned in Section~\ref{sec:relwork}, which report near-perfect results and that use an over-sampling algorithm within their pipeline, to conduct two experiments. On the one hand, we apply over-sampling before data partitioning to show that only then we are able to come close to the performance mentioned in the corresponding studies. On the other hand, we investigate the added value of tuning the over-sampling algorithm and the different hyper-parameters, when applied after data partitioning. We further clarify that in this experiment we merely want to reproduce the results of the published methodologies and do not intend to outperform previous solutions as no feature subset selection or advanced classification techniques have been applied. \\

In Table~\ref{tab:reproduction}, we report the classification technique, the over-sampling algorithm used, the reported evaluation metric and the reproduced evaluation metric when applying over-sampling before data partitioning for each of the studies. A feature set as similar as possible to the original study was used. In Table~\ref{tab:oversampling} we use these methodologies to compare the AUC of the following cases (i) when no over-sampling is applied, (ii) when the same over-sampling algorithm as in the corresponding studies is used (with default hyper-parameters) and correctly applied after data partitioning, (iii) when the hyper-parameters of the same over-sampling algorithm are tuned, and (iv) when the algorithm itself, and its hyper-parameters are tuned. In order to tune the over-sampling algorithm itself (i.e. pick the optimal one), 19 different algorithms from the smote-variants library~\cite{kovacs2019smote} are used. These algorithms were shown to achieve the best predictive performances according to an experiment with a large number of different datasets having varying properties~\cite{smote-comparison}. All results are generated using 10-fold stratified cross-validation. \\

Table~\ref{tab:reproduction} shows that we are able to rather closely approximate the reported predictive performances of the related works. Unfortunately, this could only be achieved by performing over-sampling before data partitioning. In Section~\ref{sec:oversampling}, we elaborated on how this causes leakage and results in overly optimistic performances. More realistic results of those methodologies can be found in the `Default' column of Table~\ref{tab:oversampling}. As can be seen, the discrepancy between those two is immense. It should also be noted that only AUC scores are reported in Table~\ref{tab:oversampling}, to allow for comparison between studies. While this makes the comparison for studies only reporting accuracy more difficult, we argue that accuracy is not an ideal metric to assess the predictive performance. Despite it being very comprehensible, it can give a skewed view in the context of imbalanced data. As an example, a naive model always predicting `term' on this dataset would achieve a rather high accuracy of $87.25\%$. \\

The results in Table~\ref{tab:oversampling} show a positive effect when correctly applying over-sampling (i.e., after data partitioning), as the AUC increases roughly 3 to 10\% when compared using no over-sampling. Moreover, we often notice an increase in the AUC when the hyper-parameters of the over-sampling algorithm are tuned. Tuning the number of generated minority samples seems to have the most impact, as we notice increases up to roughly 6\% in the AUC scores between the `Default' and `Tuned' column. This is in contrast with the reproduced studies, where the data was merely balanced. Additional improvements can be achieved by tuning the over-sampling algorithm itself, which is similar to the model selection phase when creating a machine learning pipeline, as the differences between the `Tuned' and `Best' columns show. It should, however, be noted that these improvements are often rather marginal. Nevertheless, from these results, we can conclude that using SMOTE, or one of its variants, with its default-hyper-parameters often results in good performances, but that different hyper-parameter configurations or other over-sampling algorithms such as NEATER or CBSO are worth investigating as well. Finally, we would like to highlight that the improvements in AUC scores achieved by the over-sampling techniques are aligned with the findings of Kov\'acs~\cite{smote-comparison}, who evaluated the same over-sampling techniques on 104 imbalanced datasets comparable to TPEHGDB in size: the average improvements achieved in terms of AUC fall in the range 4\%-10\%. Therefore, improving the AUC by a range of 40\%-50\% by simply using over-sampling techniques, as reported in some previous works, is highly unlikely.

\begin{table}
	\centering
	\small
	\begin{tabular}{|c|c|c|c||c|c|}
		\toprule Study & Classifier & Over-sampler & Metric & Report. & Reprod. \\ \midrule
		\cite{fergus2013prediction} & SVM & SMOTE & AUC & $92.0$ & $96.57$ \\
		\cite{ren2015improved} & AB & SMOTE & AUC & $98.6$ & $96.13$ \\
		\cite{idowu2015artificial} & RF & SMOTE & AUC & $94.0$ & $96.40$ \\
		\cite{ahmed2016multivariate} & SVM & ADASYN & AUC & $99.0$ & $99.13$ \\
		\cite{fergus2016advanced} & NN & SMOTE & AUC & $94.0$ & $95.63$  \\
		\cite{jager2018characterization} & QDA & ADASYN & AUC & $99.44$ & $99.28$  \\
		\cite{peng2019evaluation} & RF & ADASYN & AUC (Early) & $88.8$ & $91.72$ \\ \midrule
		\cite{hussain2015dynamic} & RF & min/max & Accuracy & $83.0$ & $85.58$ \\
		\cite{acharya2017automated} & SVM & ADASYN & Accuracy & $97.1$ & $98.46$ \\
		\cite{hoseinzadeh2018use} & SVM & ADASYN & Accuracy & $96.25$ & $93.65$ \\
		\cite{khan2019characterization} & SVM & ADASYN & Accuracy & $95.5$ & $98.65$ \\ \bottomrule
	\end{tabular}
	\caption{Comparing the AUC results (column `Report') reported in published works (reference in column `Study') with our own implementation that uses a similar feature set, the same classification technique (`Classifier') and over-sampling algorithm (`Over-sampler'), applied (incorrectly) before data partitioning (column `Reprod.'). SVM = Support Vector Machine, QDA = Quadratic Discriminant Analysis, RF = Random Forest, AB = AdaBoost, NN = Neural Network.}
	\label{tab:reproduction}
\end{table}

\begin{table}
	\centering
	\scriptsize
	\begin{tabular}{|c|c|c||c|c|c|c|}
		\toprule Study & Classif. & Over-samp. & None & Default & Tuned & Best \\ \midrule
		\cite{fergus2013prediction} & SVM & SMOTE & $49.54$ &  $60.75$ & $60.75$ & $60.85$ (NEATER~\cite{neater}) \\
		\cite{ren2015improved} & AB & SMOTE & $46.38$ & $57.43$ & $57.54$ & $58.23$ (Cluster SMOTE~\cite{cluster_SMOTE}) \\ 
		\cite{hussain2015dynamic} & RF & min/max & $42.68$ & $46.79$ & $52.89$ & $54.34$ (CBSO~\cite{cbso}) \\
		\cite{idowu2015artificial} & RF & SMOTE & $43.78$ & $46.97$ & $52.31$ & $52.31$ (SMOTE~\cite{smote}) \\
		\cite{ahmed2016multivariate} & SVM & ADASYN & $50.81$ & $54.28$ & $56.04$ & $57.47$ (NEATER~\cite{neater}) \\
		\cite{fergus2016advanced} & NN & SMOTE & $49.10$ & $50.74$ & $50.74$ & $52.11$ (LVQ SMOTE~\cite{lvq_smote}) \\
		\cite{acharya2017automated} & SVM & ADASYN & $54.80$ & $56.54$ & $57.75$ & $58.65$ (NEATER~\cite{neater}) \\
		\cite{jager2018characterization} & QDA & ADASYN & $57.51$ & $62.36$ & $63.86$ & $65.33$ (SMOTE Tomek~\cite{smote_tomeklinks_enn}) \\
		\cite{hoseinzadeh2018use} & SVM & ADASYN & $47.85$ & $49.45$ & $49.68$ & $56.03$ (CBSO~\cite{cbso}) \\
		\cite{khan2019characterization} & SVM & ADASYN & $54.72$ & $56.46$ & $57.07$ & $57.17$ (Selected SMOTE~\cite{smote_out_smote_cosine_selected_smote}) \\
		\cite{peng2019evaluation} & RF & ADASYN & $48.40$ & $51.85$ & $51.85$ & $55.05$ (Polyn. SMOTE~\cite{polynomial_fit_smote}) \\
		\bottomrule
	\end{tabular}
	\caption{The achieved AUC scores, using the methodologies of related works, but with over-sampling applied after data partitioning. We compare the AUC of (i) when no over-sampling is used (`None'), (ii) using the same over-sampling technique as the corresponding work with default hyper-parameters (`Default'), (iii) using the same over-sampling technique with tuned hyper-parameters (`Tuned'), and (iv) using the best-performing over-sampling algorithm that is optimally tuned (`Best'). SVM = Support Vector Machine, QDA = Quadratic Discriminant Analysis, RF = Random Forest, AB = AdaBoost, NN = Neural Network.}
	\label{tab:oversampling}
\end{table}

\section{Conclusion and Future Work}
In light of a significant body of recent literature reporting near-perfect results on the TPEHGDB dataset, we showed how subtle details in the methodology can introduce label leakage, which results in overly optimistic results. One of these potential pitfalls is over-sampling for data-augmentation purposes performed prior to partitioning data into training and evaluation sets. In order to investigate the actual impact of over-sampling, we re-implemented the features proposed in recent work and extracted them from the TPEHGDB dataset. We reproduced the results of 11 studies that reported near-perfect results and used an over-sampling algorithm in their pipeline. We demonstrated that we can only approximate their reported results when over-sampling is applied \textit{before} data partitioning. Next, we assessed what the actual impact of over-sampling would be in their methodologies, when applied \textit{after} data partitioning. Moreover, we investigated the added value of tuning the over-sampling algorithm and its hyper-parameters. Our results indicate that further research endeavours are required before preterm birth risk estimations based on EHG signals in a clinical setting might become useful in practice. \\

To support further research in this area, and in order to stimulate the correct use of over-sampling techniques in general, we make all code used for this study publicly available. We envision different research directions to be very promising. First, the collection of a larger dataset, with more uniform recording times and containing an important fraction of preterm cases will be required to build clinically useful models. Second, predicting entire curves of probabilities of a patient being pregnant at a certain point in time through the use of survival analysis instead of single-point predictions as in previous studies may increase the usefulness of a predictive model. Finally, deep learning could be an interesting future research approach as it is able to automatically learn representations of the EHG signals, as opposed to feature engineering.
    
\section*{Conflict of interest statement}

The authors declare no competing interests.

\section*{Acknowledgements}
Gilles Vandewiele (1S31417N) and Isabelle Dehaene (1700520N) are funded by a scholarship of FWO. This study has been performed in the context of the `Predictive health care using text analysis on unstructured data project', funded by imec, and the PRETURN (PREdiction Tool for prematUre laboR and Neonatal outcome) clinical trial (EC/2018/0609) of Ghent University Hospital. All funding bodies played no role in the creation of this study.

\section*{Code and data availability}

The code is available on Github under an open license\footnote{\url{https://github.com/GillesVandewiele/EHG-oversampling/}}. The TPEHGDB dataset is available from Physionet\footnote{\url{https://physionet.org/content/tpehgdb/1.0.1/}}.

\bibliography{bibliography}

\end{document}